\documentclass[letterpaper, 11pt, notitlepage]{article}
\usepackage{osameet3}

\usepackage[T1]{fontenc}
\usepackage{float}
\usepackage{graphicx}
\usepackage{epstopdf}
\usepackage{amssymb}
\usepackage{amsmath}
\usepackage[normalem]{ulem}
\usepackage[abs]{overpic}
\usepackage{color}
\usepackage{rotating}
\usepackage{subfigure}
\usepackage{physics}
\usepackage{bm}
\usepackage{enumerate}
\usepackage[utf8]{inputenc}
\usepackage{booktabs}
\usepackage{placeins}

\usepackage{lineno}
\usepackage{setspace}

\makeatletter

\usepackage[style=phys,biblabel=superscript,chaptertitle=false]{biblatex}
\usepackage[english]{babel}
\usepackage{blindtext,tikz}
\usetikzlibrary{calc}

\usepackage{siunitx}

\newcommand{\etal}[0]{\textit{et al.}}
\newcommand{\insitu}[0]{\textit{in situ}}
\renewcommand{\figurename}{Fig.}
\begin{document}

\title{On-chip sampling of optical fields with attosecond resolution}
\date{\today}
\author{\small{Mina R. Bionta$^{1,\dagger,*}$, Felix Ritzkowsky$^{2,\dagger,*}$, Marco Turchetti$^{1,\dagger}$, Yujia Yang$^1$, Dario Cattozzo Mor$^1$, William P. Putnam$^3$, Franz X. Kärtner$^2$,  Karl K. Berggren$^1$, and Phillip~D.~Keathley$^{1,*}$}}
\address{$^1$Research Laboratory of Electronics, Massachusetts Institute of Technology, 77 Massachusetts Avenue, Cambridge, MA 02139, USA\\
$^2$Deutsches Elektronen Synchrotron (DESY) \& Center for Free-Electron Laser Science, Notkestra\ss e 85, 22607 Hamburg, Germany\\
$^3$Department of Electrical and Computer Engineering, University of California, Davis, 1 Shields Ave, Davis, CA 95616, USA\\
$^\dagger$These authors contributed equally to this work.
}
\email{$^*$e-mail: mbionta@mit.edu; felix.ritzkowsky@desy.de; pdkeat2@mit.edu}

\maketitle

\vspace{10pt}
\vspace{10pt}
\textbf{Time-domain sampling of arbitrary electric fields with sub-cycle resolution enables a complete time-frequency analysis of a system's response to electromagnetic illumination.
This time-frequency picture provides access to dynamic information that is not provided by absorption spectra alone, and has been instrumental in improving our understanding of ultrafast light-matter interactions in solids that give rise to nonlinear phenomena~\supercite{hassan_optical_2016, schubert_sub-cycle_2014, Hohenleutner2015, bonvalet_femtosecond_1996}.
Furthermore, it has recently been shown through measurements in the infrared that nonlinear, sub-cycle, optical-field sampling offers significant improvements with regard to molecular sensitivity and limits of detection compared to traditional spectroscopic methods for the characterization of biological systems~\supercite{Pupeza2020}. However, despite the many scientific and technological motivations, time-domain, sub-cycle, optical-field sampling systems operating in the visible to near-infrared spectral regions are seldom accessible, requiring large driving pulse energies, and accordingly, large laser amplifier systems, bulky apparatuses, and vacuum environments \supercite{Pupeza2020, Keiber2016, Kim2013, Park2018, Wirth2011, Krausz2014}.
Here, we demonstrate an all-on-chip, optoelectronic device capable of sampling arbitrary, low-energy, near-infrared waveforms under ambient conditions.
Our solid-state integrated detector uses optical-field-driven electron emission from resonant nanoantennas to achieve petahertz-level switching speeds by generating on-chip attosecond electron bursts \supercite{Rybka2016, Ludwig2020, Keathley2019, Putnam2017}.
These bursts are then used to probe the electric field of weak optical transients.
We demonstrated our devices by sampling the electric field of a $\sim$5~\si{\femto\joule} (6.4~MV~m$^{-1}$), broadband near-infrared ultrafast laser pulse using a $\sim$50~\si{\pico\joule} (0.64~GV~m$^{-1}$) near-infrared driving pulse.
Our sampling measurements recovered the weak optical transient as well as localized plasmonic dynamics of the emitting nanoantennas \insitu.
This field-sampling device -- with its compact footprint and low pulse-energy requirements -- offers opportunities in a variety of applications~\supercite{Hohenleutner2015, Pupeza2020, Dombi2020}, including: broadband time-domain spectroscopy in the molecular fingerprint region, time-domain analysis of nonlinear phenomena, and detailed studies of strong-field light-matter interactions.}

Complimentary time-frequency analysis enabled by time-domain sampling is critical to the understanding and design of electronic systems, and such studies have revolutionized spectroscopy in the terahertz spectral region \supercite{Neu2018}.
Commercial THz time-domain spectroscopy systems are now readily available \supercite{ThorlabsTHz} and often used for industrial applications such as chemical and material analysis.
Sub-cycle field sampling in the THz regime has also been instrumental to many fundamental scientific investigations, including the tracing of electron wavepacket dynamics in quantum wells~\supercite{bonvalet_femtosecond_1996}, the investigation of dynamic Bloch oscillations in semiconductor systems~\supercite{schubert_sub-cycle_2014}, and the observation and characterization of quantum vacuum fluctuations~\supercite{riek_direct_2015}.

Optical-field sampling in the visible to near-infrared (near-IR) spectral regions would provide great benefit to both science and industry.
For example, attosecond streaking spectroscopy has been used to study the role of optical-field-controlled coherent electron dynamics in the control of chemical reaction pathways~\supercite{lepine_attosecond_2014} and the investigation of petahertz-level electrical currents in solid-state systems~\supercite{Schultze2013, schiffrin_optical-field-induced_2013}.
It was also recently shown that sub-cycle field sampling of the free-induction decays of biological systems can provide an order of magnitude reduction in the limits of detection and improved molecular sensitivity compared to traditional frequency-domain spectroscopic methods~\supercite{Pupeza2020}.
Despite these compelling results, scaling such techniques into the near-IR and visible spectral regions has remained challenging.
Manipulation of short electron wave packets \supercite{Park2018, Cho2019} and attosecond streaking in the visible to near-IR spectral regions \supercite{Itatani2002, Kienberger2004, Schotz2017, Wyatt16} have proven to be viable paths towards direct optical-field sampling in the time-domain.
However, these techniques require high-energy optical sources and complicated optical apparatus, with no compact and integratable sampling technology with the bandwidth and field sensitivity required for real-world applications of interest.

To address this lack of compact and integratable tools for optical-field sampling in the visible to near-IR, we have developed and demonstrated an on-chip, time-domain, sampling technique for measuring arbitrary electric fields of few-\si{\femto\joule} optical pulses in ambient conditions.
The enhanced local electric field surrounding plasmonic nanostructures has been used to generate strong electric fields in nanometer sized volumes creating a new regime for exploring attosecond science \supercite{Dombi2020, Kruger2018, Schoetz2019, Ciappina2017, Stockman2018}.
Our work leverages the sub-cycle optical-field emission from plasmonic nanoantennas to achieve petahertz-level sampling bandwidths using only picojoules of energy \supercite{Ludwig2020,Putnam2017,Keathley2019, Kruger2011}. Furthermore, by electrically connecting the nanoantenna arrays via nanoscale wires, the field samplers we demonstrate here are amenable to large-scale electronic integration~\supercite{Yang2019, Ludwig2020b}.
Beyond demonstrating the feasibility of sub-cycle field sampling of petahertz-scale frequencies, our results also reveal \insitu~dynamical properties of the interaction of the driving optical-field waveform with the plasmonic nanoantennas.
This work will enable the development of new tools for optical metrology that will complement traditional spectroscopic methods and unravel linear and nonlinear light-matter interactions as they occur at their natural time and length scales.

\begin{figure}[h]
    \centering
    \includegraphics{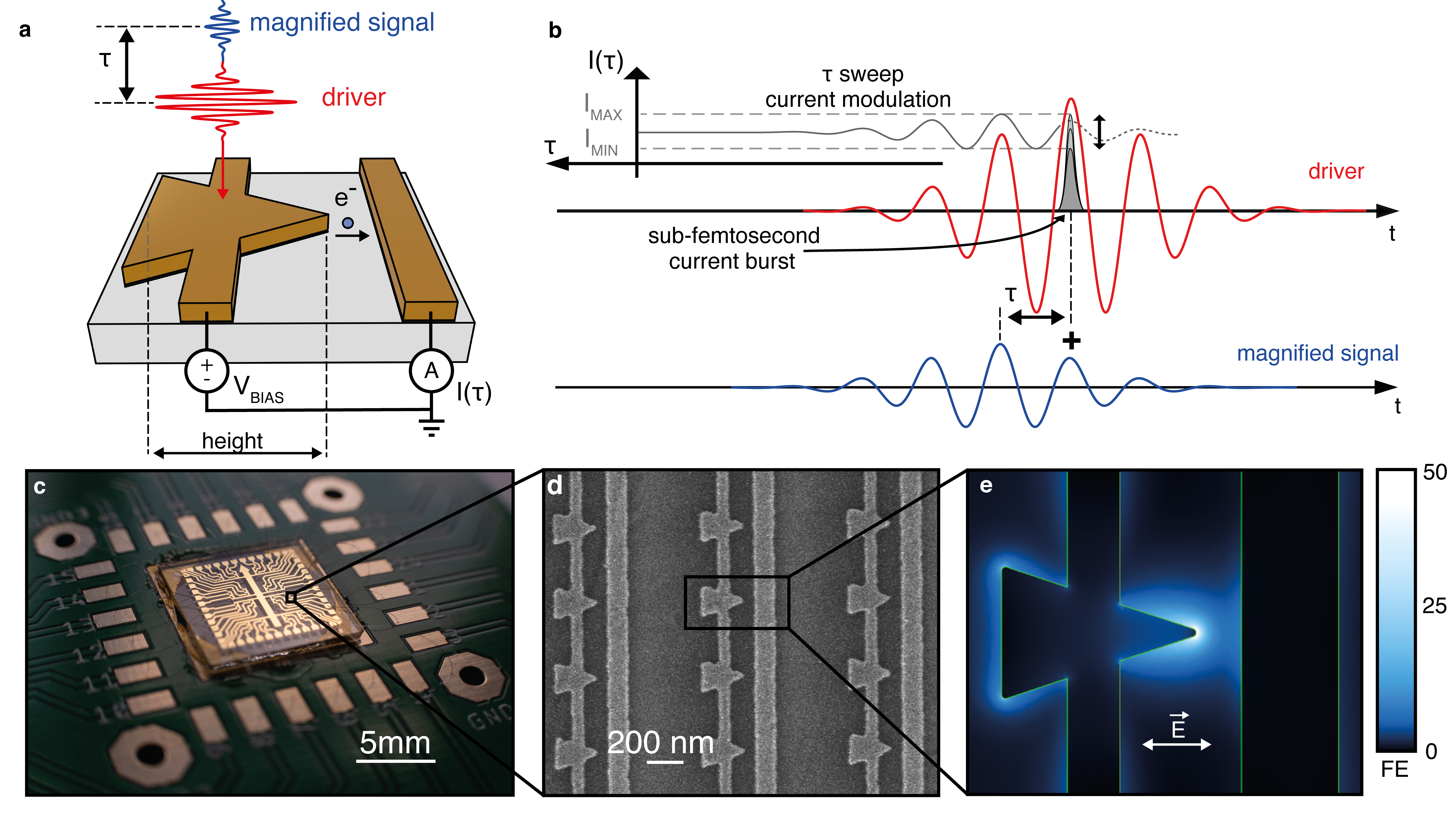}
    \caption{\textbf{Device overview.}
    \textbf{a,} Schematic of the device.
    \textbf{b,} Depiction of the optical-field sampling process.
    Attosecond electron bursts are driven from an electrically-connected gold nanoantenna (see \textbf{a}) by a strong optical waveform (driver, red), collected by an adjacent gold wire, then measured using an external current detector ($I(\tau)$ is the measured time-averaged current; see Methods).
    The weak signal waveform (blue), with a peak intensity of $1\times10^{-4}$ that of the driver pulses, modulates the average photocurrent generated by the driver pulse, $I(\tau)$, as a function of delay, $\tau$ (grey).
    The amplitude of the signal waveform is artificially magnified in \textbf{a} and \textbf{b} for visibility, but would in reality be roughly two orders of magnitude smaller than the driver waveform.
    \textbf{c,} Photograph of the nanocircuit embedded on printed circuit board.
    \textbf{d,} Scanning electron micrograph of the device.
    \textbf{e,} Simulated electric field enhancement around a nanoantenna. The maximum field enhancement is $\sim$35. FE: field enhancement factor. E: arrow indicates polarization of the incident electric field.
    \label{fig:overview}}
\end{figure}

Our device is depicted in Fig.~\ref{fig:overview}a. It consists of an electrically-connected plasmonic gold nanoantenna that functions as the electron source (cathode), a gold nanowire as an anode separated by a 50~nm air gap, and an external current detector.
A photograph of the nanocircuit integrated onto a printed circuit board is shown in Fig. \ref{fig:overview}c.
Devices were connected in parallel via nanowires and simultaneously excited to improve signal strength.
A scanning electron micrograph (SEM) of the fabricated devices is shown in Fig.~\ref{fig:overview}d.

When the strong driving pulse $E_{\mathrm{D}}(t)$ illuminates the nanoantenna/wire junction, a large local electric field $E^{(\mathrm{L})}_{D}(t)$ is generated at the nanoantenna tip as shown in  Fig.~\ref{fig:overview}e. The incident electric field is related to the local electric field by the transfer function of the nanoantenna $\tilde{H}_\mathrm{Pl.}(\omega)$ by the relationship $E^{(\mathrm{L})}_\mathrm{D}(t)=\mathcal{F}
^{-1}\left(\tilde{H}_\mathrm{Pl.}(\omega)\cdot \tilde{E}_{\mathrm{D}}(\omega)\right)$, where $\mathcal{F}
^{-1}$ is the inverse Fourier transform, and tildes indicate the Fourier domain. Due to the combined effect of the localized surface plasmon polariton \supercite{Stockman2011} in the antenna and the geometric field enhancement resulting from the sharp radius of curvature \supercite{Gomer1961}, the locally-enhanced field exceeds the incident electric field of the driver pulse by a factor of $\sim$35.
The weak incident signal field $E_\mathrm{S}(t)$ then creates a weak local signal field $E^\mathrm{(L)}_\mathrm{S}(t)$ that modulates the average photocurrent, $I(\tau)$, from which the local signal field can be determined as a function of delay, $\tau$, between the two pulses (Fig.~\ref{fig:overview}b).
The local signal field itself is too weak to drive photoemission, and thus it only modulates the electron burst(s) emitted by the local driver field.

If sufficiently strong, the local driving electric field at the antenna tip $E^{(\mathrm{L})}_\mathrm{D}(t)$ significantly bends the surface potential, resulting in optical-field-driven tunneling of electrons at the metal-vacuum interface once every cycle~\supercite{Putnam2017, Rybka2016, Yang2019, Keathley2019}.
The instantaneous emission rate $\Gamma$ approaches the static tunneling emission rate defined by Fowler and Nordheim, $
\Gamma\left(E\right) \propto E^2 \cdot \exp^{-\frac{F_t}{\abs{E}}}$~\cite{Fowler1928} as described in Ref.~\cite{Bunkin1965} and Ref.~\cite{Yalunin2011}.
The characteristic tunneling field strength $F_t = \SI{78.7}{\volt\per\nano\meter}$ is dependent on the work function of the metal, approximately \SI{5.1}{\electronvolt} for gold. Due to the strong nonlinearity of the emission process, the electron bursts generated in the device are deeply sub-cycle and on the order of several hundred attoseconds for the case of near-IR fields \supercite{Ludwig2020,Keathley2019}.

For calculating impact of the weak signal field on the total emission, a linearized small-signal model can be used.
We consider the addition of the weak local signal $E_\mathrm{S}^{(\mathrm{L})}(t)$ as a function of delay $\tau$ relative to the strong driving field $E_\mathrm{D}^{(\mathrm{L})}(t-\tau)$ as shown in Fig.~\ref{fig:overview}b.
As demonstrated by Cho~\etal~\supercite{Cho2019}, a short optical driving pulse in combination with a highly-nonlinear, sub-cycle emission process allows for field-resolved sampling of the signal pulse.
In our case, the detected current as a function of delay $I(\tau)$ is the time-average of the nonlinear emission rate $\Gamma$ driven by the sum of the driver field $E_\mathrm{S}^{(\mathrm{L})}(t)$ and the small-amplitude signal field $E_\mathrm{D}^{(\mathrm{L})}(t-\tau)$,
\begin{equation}
    I(\tau) \propto \int_{-\frac{T_\mathrm{Rep.}}{2}}^{\frac{T_\mathrm{Rep.}}{2}} \Gamma\left( E_\mathrm{D}^{(\mathrm{L})}(t-\tau) +  E_\mathrm{S}^{(\mathrm{L})}(t)\right)dt\mbox{,}
    \label{eq:correlation}
\end{equation}
where $T_\mathrm{Rep.}$ is the time between consecutive optical pulses.  Given that $E_\mathrm{S}^{(\mathrm{L})}(t)$ is sufficiently small, we can Taylor-expand $\Gamma$ around the local driver field $E_\mathrm{D}^{(\mathrm{L})}(t-\tau)$ to the first order.
This enables the linearization of the measured emission $I(\tau)$ with respect to the signal $E_\mathrm{S}^{(\mathrm{L})}(t)$:

\begin{equation}
    I(\tau) \propto \int_{-\frac{T_\mathrm{Rep.}}{2}}^{\frac{T_\mathrm{Rep.}}{2}} \left( \Gamma \bigl( E_\mathrm{D}^{(\mathrm{L})}(t-\tau)\bigr) + \left.\dv{\Gamma}{E}\right\vert_{ E_\mathrm{D}^{(\mathrm{L})}(t-\tau)} \cdot E_\mathrm{S}^{(\mathrm{L})}(t)\right) dt.
    \label{eq:taylor}
\end{equation}

\begin{figure}[h]
    \centering
    \includegraphics[width=\linewidth]{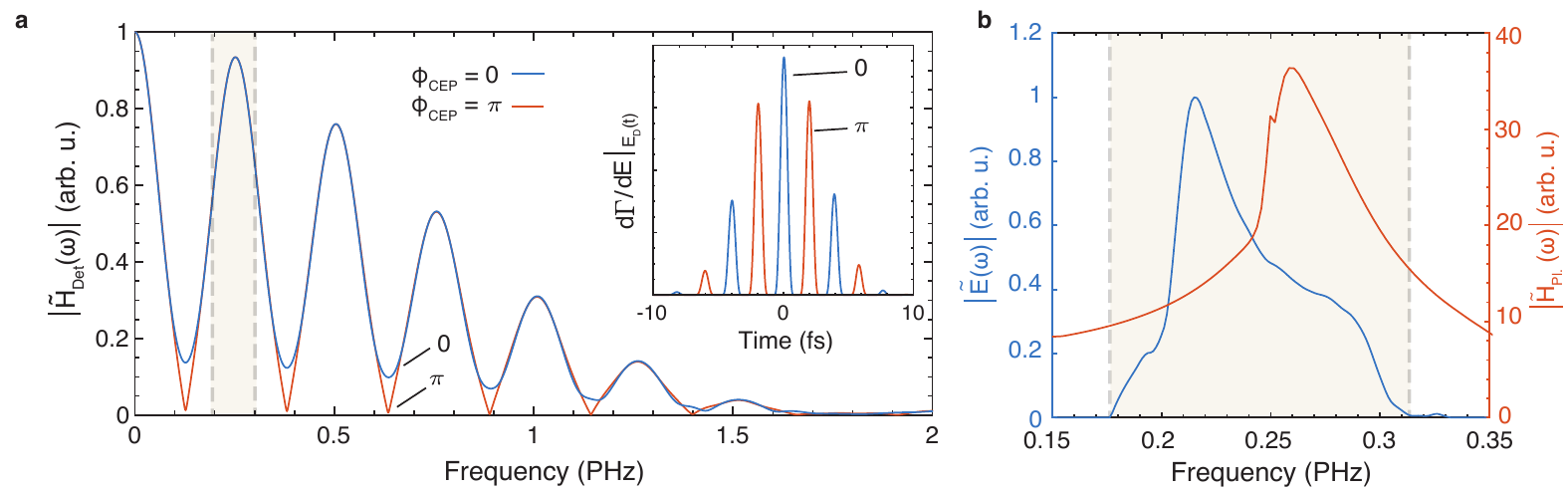}
    \caption{\small\textbf{Theoretical sampling bandwidth.} \textbf{a,} Calculation of the accessible sampling bandwidth $\tilde{H}_{\mathrm{Det}}(\omega)$ as dictated by the Fourier transform of $\left.\dv{\Gamma}{E}\right\vert_{E_\mathrm{D}^{(\mathrm{L})}(t)}$ for the carrier-envelope phases  $\Phi_{\mathrm{CEP}}=0, \pi$ of the driver pulse (blue and red respectively). (Inset) The time-domain picture of $\left.\dv{\Gamma}{E}\right\vert_{E_\mathrm{D}^{(\mathrm{L})}(t)} $ corresponding to the bandwidth shown in \textbf{a}. \textbf{b,} The spectral amplitude of the driving pulse (blue) and the plasmonic nanoantenna transfer function $\left\vert \tilde{H}_\mathrm{Pl.}(\omega)\right\vert$ (red). The shaded area in \textbf{a} and \textbf{b} indicates the spectral region occupied by the driving pulse. \label{fig:theory}}
\end{figure}

The second term in Eq.~\ref{eq:taylor} is a cross-correlation between $\left.\dv{\Gamma}{E}\right\vert_{E_\mathrm{D}^{(\mathrm{L})}(t-\tau)}$ and $E_\mathrm{S}^{(\mathrm{L})}(t)$, and denoted as $I_{\mathrm{CC}}(\tau)$. Due to the nonlinearity of the emission process, the central most portion of the driving waveform dominates the measured time-integrated current, and acts a sub-cycle gate limiting interaction with the signal field (Fig.~\ref{fig:theory}a inset). This fact becomes more evident when taking the Fourier transform of $I_{\mathrm{CC}}(\tau)$, which simplifies to the following expression:
\begin{equation}
    \tilde{I}_{\mathrm{CC}}(\omega)\propto \mathcal{F}\left(\left.\dv{\Gamma}{E}\right\vert_{ E_\mathrm{D}^{(\mathrm{L})}(t)}\right) ^*\cdot \tilde{E}_\mathrm{S}^{(\mathrm{L})}(\omega),
    \label{eq:fourierDomain}
\end{equation}
where $\mathcal{F}\Bigl(\left.\dv{\Gamma}{E}\right\vert_{ E_\mathrm{D}^{(\mathrm{L})}(t)}\Bigr) ^*$ is the complex spectrum shown in Fig.~\ref{fig:theory}a and is denoted as $\tilde{H}_{\mathrm{Det}}(\omega)$.
This function $\tilde{I}_\mathrm{CC}(\omega)$ describes the full sampling response to the weak signal and is connected to the measured cross-correlation in the time-domain $I_\mathrm{CC}(t)$ by a Fourier transform.

The accessible bandwidth is found through the multiplication of the initial small signal $\tilde{E}_\mathrm{S}^{(\mathrm{L})}(\omega)$ with $\tilde{H}_{\mathrm{Det}}(\omega)$.  Due to the highly-nonlinear sub-cycle response of the emission rate on the driving electric field, $\tilde{H}_{\mathrm{Det}}(\omega)$ spans several octaves from DC to more than $1~\mathrm{PHz}$ (see  Fig.~\ref{fig:theory}a). The resultant small-signal gain  enhances the response of the system to weak signals of interest $\tilde{E}_\mathrm{S}^{(\mathrm{L})}(\omega)$ that would not be able to generate detectable electron emission signals on their own.
We should also note the periodic structure in the plot of $\tilde{H}_{\mathrm{Det}}(\omega)$ shown in Fig.~\ref{fig:theory}a.  This structure is due to the presence of regularly spaced electron pulses in the time domain which result in the modulation of the amplitude of $\tilde{H}_{\mathrm{Det}}(\omega)$ (see inset of Fig.~\ref{fig:theory}a).
These pulses are spaced by one cycle of the driving laser, and change in number and strength depending on the carrier-envelope phase (CEP).
However, for few-cycle pulses, such as those used in this experiment, the response remains relatively flat over the bandwidth of the driving pulse, and is only minimally affected by the modulation.
More detailed discussion of CEP and pulse duration effects can be found in the Supplementary Information sections 2 and 3.

There are several practical considerations for sampling either the local or incident signal fields.
The impulse-response function of the antenna, specifically the resonant plasmonic contribution, redistributes frequency components as shown by the field enhancement as a function of frequency in Fig. \ref{fig:theory}b.
Considering the actual sampling process, the limiting factors are the satellite electron pulses, which, if too pronounced, can cause stronger modulation of the sampling bandwidth, and the breakdown of the theoretical approximations mentioned above (see Supplementary Information section~2).
If the intensity of the probed weak signal pulse is comparatively strong, approximately three orders of magnitude below the driving pulse and higher, it will result in nonlinear distortions causing higher order terms in the cross-correlation to become significant.
Another important consideration is the work function of the emitter.
If the signal pulse reaches photon energies higher than the work function, linear photoemission due to single-photon absorption will cause a substantial background current.
This places an upper frequency limit of a gold device near 1~PHz.

To experimentally verify the device performance, a CEP-stable, 78 MHz Er:fiber-based laser source was used~\supercite{Putnam2019}.
The pulses were spectrally broadened in a highly non-linear fiber to create a pulse duration down to $\sim$ \SI{10}{\femto\second} full-width at half-maximum (FWHM) ($\sim$2.5 cycles) at a central wavelength of $\sim$\SI{1170}{\nano\meter}.
Spectral phase characterization of the laser source was performed using Two-Dimensional Spectral Shearing
Interferometry (2DSI)~\supercite{Birge2006} and can be found in Supplmenentary Information section 6.
These pulses were locked to a fixed CEP value for all measurements.
A dispersion balanced Mach-Zehnder interferometer was used to generate pairs of strong driver and weak signal pulses with a variable delay for the experiment.
The driver and signal pulse energies (fields) were measured to be approximately \SI{50}{\pico\joule} (0.64~GV~m$^{-1}$ at focus) and $\sim$\SI{5}{\femto\joule} (6.4~MV~m$^{-1}$ at focus) respectively.
The two pulses were focused to a spot-size of $\SI{2.25}{\micro\meter}\times\SI{4.1}{\micro\meter}$~FWHM, illuminating 10-15 nanoantennas at a time.
The pulses were linearly polarized along the height axis of the nanoantennas (Fig.~\ref{fig:overview}a).
Attosecond electron bursts were primarily generated in the nanocircuit by the \SI{50}{\pico\joule} driver pulse, with the emission modulated by the $\sim$\SI{5}{\femto\joule} signal pulse as the delay between the two pulses was scanned.
The signal pulses, with an intensity of $1\times10^{-4}$ that of the drive pulses, were much too weak to drive electron emission on their own.
The photocurrent was then detected using a transimpedance amplifier in conjunction with lock-in detection.
We emphasize that the experiment was performed in ambient conditions (\textit{i.e.} in air and at room temperature).
This ability to function in ambient conditions is enabled by the small gap size between the nanoantenna and the collecting wire \supercite{Putnam2017, Rybka2016}.
A schematic of the experiment is shown in Fig.~\ref{fig:overview}a and Supplementary Fig.~S1 with further details found in the Methods section and Supplementary Information Sec.~1.

\begin{figure}[h]
    \centering
    \includegraphics{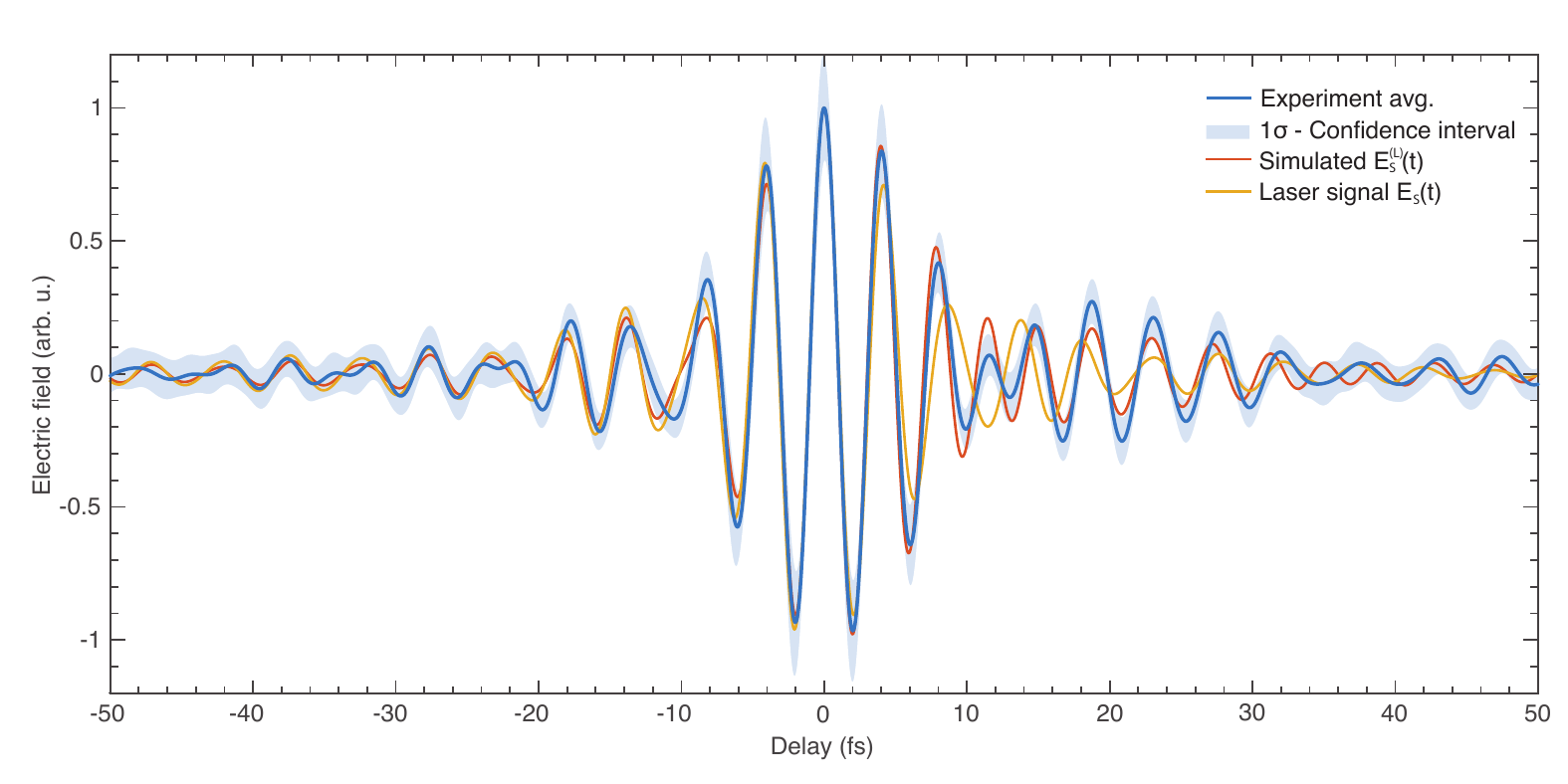}
    \caption{\small\textbf{Experimental field sampling results and analysis.}
    Time-domain results for devices with a 240~nm height comparing measured (blue) and simulated near-fields (red) to the calculated incident laser signal (yellow).
    Here, negative delays indicate the driver pulse arrives before the signal pulse. The 1$\sigma$-confidence interval is shown as a blue shaded ribbon centered at the average value (blue solid line) retrieved from 60 scans.
    The plasmonic resonance of the antenna results in a dephasing in the time-domain between $E_\mathrm{S}(t)$ and $E_\mathrm{S}^{(\mathrm{L})}(t)$ as observed around \SI{12}{fs}.
    \label{fig:Data}}
\end{figure}
\begin{figure}[h]
    \centering
    \includegraphics[width=0.7\textwidth]{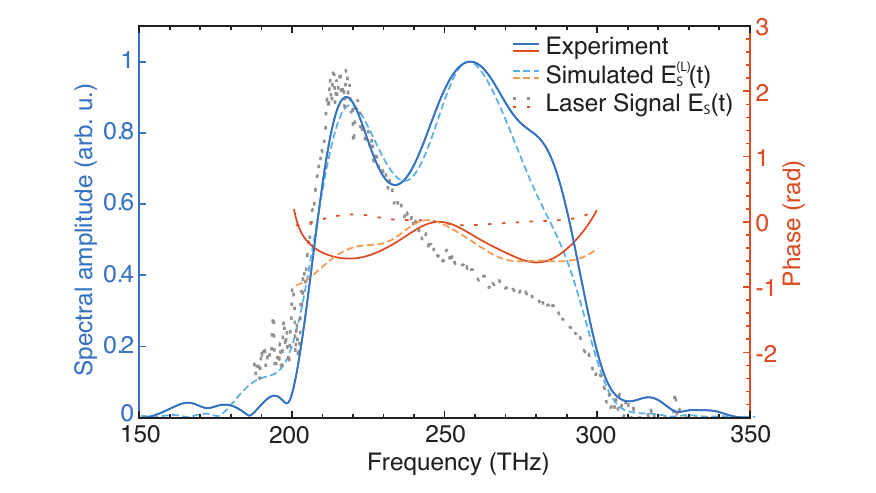}
    \caption{\small\textbf{Frequency-domain of the experimental field sampling results}
    Frequency-domain comparison of measured (solid) and simulated (dashed) near-fields for devices with a 240~nm height to the calculated incident laser signal (dotted). This on-resonant 240~nm device shows two peaks present in the cross-correlation data, one corresponding to the output laser spectrum (at 218~THz) and the other to the plasmonic enhancement of the antenna (at 257~THz).
    \label{fig:DataSpectrum}}
\end{figure}

Fig~\ref{fig:Data} presents the measured cross-correlation (blue trace) for the tested antennas with a 240~nm height (from antenna base to tip, see Fig.~\ref{fig:overview}a), and compares them with the computed simulated antenna response $E_\mathrm{S}^{(\mathrm{L})}(t)$ (red trace) and the calculated laser signal $E_\mathrm{S}(t)$ (yellow trace).
The calculated laser signal $E_\mathrm{S}(t)$ was found by applying the measured spectral phase (see Supplmenentary Information section \ref{sec:2DSI}) of the laser output to the measured intensity spectrum before converting back to the time domain.
The measured trace (blue) shows significant deviations from the calculated laser signal $E_\mathrm{S}(t)$ (yellow), especially in the pedestal from 5~fs to 20~fs. However, this measured pulse shape in the time-domain (blue trace) is almost identical to the simulated antenna response, $E_\mathrm{S}^{(\mathrm{L})}(t)$ (red trace), both with a 180$^{\circ}$ dephasing near \SI{12}{\femto\second} with respect to the incident laser signal $E_\mathrm{S}(t)$ (yellow trace). Similar dephasing dynamics have been investigated by others in both nanoantenna and extended nanotip structures~\supercite{sun_direct_2013, hanke_efficient_2009, anderson_few-femtosecond_2010} and are a hallmark of the resonant electron dynamics excited within the nanoantennas. The $1\sigma$-confidence interval (that is the interval ranging between plus and minus one standard deviation from the mean value) was calculated over all 60 scans and is shown as the light blue shaded region in Fig.~\ref{fig:Data}. One standard deviation is $\approx \SI{10}{\%}$ of the peak amplitude. Considering the estimated peak field of around \SI{6.4}{MV m^{-1}} for the incident laser signal, the detection floor is estimated around \SI{600}{kV m^{-1}}.

In the frequency-domain (Fig~\ref{fig:DataSpectrum}), two prominent maxima are visible in the Fourier transform of the measured data (blue solid trace).
These maxima are also exhibited in the simulation of the antenna response (light blue dashed trace), but only one (at $218$~THz) is observed in the measured laser spectrum (grey dotted trace).
The second peak (at $257$~Thz) is due to the plasmonic response of the antenna $\tilde{H}_\mathrm{Pl.}(\omega)$ which must be incorporated when calculating the electric near-field.
The peak in the spectral phase of the measured data (red solid trace) is due to the plasmonic resonance of the antenna and closely matches the simulation of the antenna response (orange dashed trace). This spectral analysis further supports our conclusion that the observed discrepancies and dephasing between the incident laser signal $E_\mathrm{S}(t)$ and the measured pulse in the time-domain (Fig.~\ref{fig:Data}) arise due to the resonant response of the nanoantenna. Similar experimental results and analysis for 200~nm devices can be found in Supplementary Information section~4.

We attribute the minor discrepancies between the simulated and experimental data to the multiplexed nature of our current detection and minor uncertainties in the fabrication process which were not accounted for in our models.
As we illuminated 10-15 nanoantennas at a time, the measurements we show are an averaged trace, with all antennas contributing simultaneously to the detected current.
This averaging causes the detected resonance shape to be a superposition of all antenna resonances.
Another possible cause of slight discrepancy is the high malleability of gold, which can allow the antennas to reshape under intense radiation, thereby creating a geometry that differs from the original shape just after fabrication~\supercite{Yang2019}.

Our detection scheme can be directly compared to hetero- and homodyne methods that are often used in techniques such as frequency-comb spectroscopy~\supercite{Picque2019, coddington_coherent_2008, Bjork2016}. In fact, the only important difference between hetero/homodyne methods and our method is the use of an energy detector in the conventional approach, as opposed to our highly nonlinear nanoantenna detector.
Energy detectors only allow for a narrow detection bandwidth that is confined to the amplitude spectrum of the local oscillator (\textit{i.e.} the driver pulse), corresponding to the shaded region in Fig.~\ref{fig:theory}.
Unlike energy detectors, the broadband response of the nonlinear nanoantenna detectors could enable simultaneous tracking of linear and nonlinear light-matter interaction dynamics.
Using few-cycle visible and near-infrared driver pulses to sample weaker, phase-locked mid-infrared transients would not require MIR detectors, with their limited capabilities, that are slow or that require cryogenic cooling. The reduced pulse energy requirements and compact form-factor of on-chip nanoantenna detectors like those presented in this work could thus be used to enhance the performance of emerging frequency-comb spectroscopy systems~\supercite{kowligy_infrared_2019}.

While other direct time-domain optical sampling techniques for visible and near-infrared optical pulses currently exist, such as time-domain observation of an electric field (TIPTOE) \supercite{Park2018, Cho2019}, and attosecond streaking \supercite{Krausz2014, Wirth2011, Itatani2002, Kienberger2004}, they require \si{\micro\joule}- to \si{\milli\joule}-level pulse energies, bulky apparatus, and/or vacuum enclosures.
By providing a compact, chip-scale platform that enables sub-cycle, field-sensitive detection of sub- to few-\si{\femto\joule} optical waveforms in ambient conditions, devices similar to those discussed in this work could find applications such as phase-resolved spectroscopy and imaging, and could have an impact in a variety of fields such as biology, medicine, food-safety, gas sensing, and drug discovery.
In particular, due to their compact footprint and \si{\pico\joule}-level energy requirements, such detectors could be used to enhance the performance and operating bandwidth of frequency comb spectroscopy systems.
Furthermore, we believe that such on-chip petahertz field-sampling devices will enable fundamental investigations such as the time-domain characterization of attosecond electron dynamics and optical-field-driven nonlinear phenomena in light-matter interactions.

\section*{Methods}
\subsection*{Experimental Methods}
The nanodevices were illuminated by a few-cycle, supercontinuum-based \supercite{Sell2009}, CEP-stablilized fiber laser source \supercite{Putnam2019}.
The source has a central wavelength of $\sim$1170~nm, with a pulse duration of $\sim$10~fs FWHM ($\sim$2.5 cycles), and repetition rate of 78~MHz.
The supercontinuum was generated from a highly non-linear gemanosilicate fiber pumped by a Er:fiber-based laser oscillator and Er-doped fiber amplifier (EDFA) system and compressed with a SF10 prism compressor.
The CEP was locked to a fixed CEP value for all measurements taken.
Pulse characterization of the laser source was performed by 2DSI~\supercite{Birge2006} and can be found in Supplmenentary Information section 6.
The spectrum of the laser source was measured with a fiber-coupled optical spectrum analyzer (Ando Electric Co., Ltd.).
More details about the supercontinuum source can be found in Ref.~\cite{Putnam2019}.

A dispersion-balanced Mach-Zehnder interferometer was used to generate the pulse pairs for the experiment.
An Inconel reflective neutral density (ND) filter of optical density (OD) 4 on a 2~mm thick BK7 substrate (Thorlabs) was placed in one arm and used to generate a weak signal pulse with pulse energy of $\sim$5~fJ.
An optical chopper was placed in this weak arm for lock-in amplification and detection.
The strong, driver arm had a pulse energy of $\sim$50~pJ.
A corresponding 2~mm thick BK7 window was placed in the driver arm to balance the dispersion between arms.
The added chirp from the glass was precompensated using the prism compressor.
The delay between the two pulses was controlled with a home built \SI{15}{\micro\meter} piezo stage.
A chopper was placed in the weak arm to modulate the signal for lock-in amplification.
A schematic of experimental setup can be found in Supplmenentary Information Sec.~1.

The pulses were focused onto the chip using a Cassegrain reflector to a spot-size of \SI{2.25}{\micro\meter}~$\times$~\SI{4.1}{\micro\meter}~FWHM.
This spot-size allowed for illumination of 10-15 nanoantennas at a time.
The polarization of the pulses was parallel to the nanoantenna height axis.
A bias voltage of \SI{3}{V} was applied across the \SI{50}{nm} device gap.
The emitted current was collected and amplified by a transimpedance amplifier (FEMTO Messtechnik GmbH) in conjunction with a lock-in amplifier (Stanford Research Systems), with a modulation of 200~Hz of the optical chopper.

For each data set, 60 scans of 10 second acquisition time over the \SI{100}{fs} time window were performed. Post-processing was done in Matlab.
Each data set was Fourier transformed and windowed from \SI{150}{THz} to \SI{350}{THz} with a tukey-window  steepness of $\alpha=0.2$. The resulting output was averaged in the time-domain.

\subsection*{Device Fabrication}
We used a fabrication process based on that described in Ref.~\cite{Yang2019}.
The data presented in this work comes from devices fabricated on two different chips. The devices were fabricated on BK7 substrates. The patterning was performed using an electron beam lithography process with PMMA A2 resist (Microchem), a writing current of 2~nA, a dose of \SI{5000}{\micro\coulomb\per\centi\meter^2}, and an electron beam energy of \SI{125}{\kilo\electronvolt}. To avoid charging, an Electra92 layer was spin-coated on top of the PMMA at 2 krpm and baked for 2 min at 90~$^{\circ}$C. Since these are large arrays, a proximity effect correction step was also included when designing the layout. After exposure, the resist was cold-developed in a 3:1 isopropyl alcohol to methyl isobutyl ketone solution for 60~s at 0~$^{\circ}$C. Then, a 2~nm adhesion layer followed by 20~nm of Au were deposited using electron beam evaporation. As adhesion layer Ti was used for the 240 nm and Cr for the 200 nm antennas chips. Subsequently a liftoff process in a 65~$^{\circ}$C bath of n-methylpyrrolidone (NMP) (Microchem) was used to release the structures. Finally, we used a photolithography procedure to fabricate the contact pads for external electrical connections.

\subsection*{Electromagnetic Simulations}
The optical response of the plasmonic nanoantennas was simulated in a finite-element-method electromagnetic solver (COMSOL Multiphysics). The nanoantenna geometry was extracted from SEM images. The refractive index of gold was taken from Ref.~\cite{johnson1972optical}, and the refractive index of the glass substrate was fixed at 1.5 with negligible dispersion in the simulation spectral range. To simulate nanoantenna arrays, periodic boundary conditions were used. The normally incident plane wave was polarized along the nanotriangle axis (perpendicular to the nanowire). Perfectly matched layers were used to avoid spurious reflections at the simulation domain boundaries. The complex field response $\tilde{H}_{\mathrm{Pl.}}(\omega)= \tilde{E}^{(L)}(\omega)/\tilde{E}(\omega)$ was evaluated as a function of frequency. The field enhancement was defined as the ratio of the near-field at the nanotriangle tip to the incident optical field.

\section*{Data and Code Availability}
The data and code that support the plots within this paper and other findings of this study are available from the corresponding authors upon reasonable request.

\printbibliography

\section*{Acknowledgements}

This material is based upon work supported by the Air Force Office of Scientific Research under award numbers FA9550-19-1-0065 and FA9550-18-1-0436. F.X.K. acknowledges support by the European Research Council under the European Union’s Seventh Framework Programme (FP7/2007–2013) through the Synergy Grant ‘Frontiers in Attosecond X-ray Science: Imaging and Spectroscopy’ (AXSIS) (609920) and by the Cluster of Excellence ‘CUI: Advanced Imaging of Matter’ of the Deutsche Forschungsgemeinschaft (DFG) - EXC 2056 - project ID 390715994. This work was also partially supported by a seed grant provided by SENSE.nano, a center of excellence powered by MIT.nano, as well as the PIER Hamburg – MIT Program.  We thank Marco Colangelo and John Simonaitis for their scientific discussion and edits to the manuscript. We thank Navid Abedzadeh for taking photos of the chip.

\section*{Author Contributions}

F.R., M.R.B. and P.D.K., conceived the experiments. Y.Y. and D.C.M. simulated the optical response of the devices. M.T. fabricated the devices. M.R.B., F.R., and M.T. performed the experiments with assistance from P.D.K. F.R. derived the theory and simulated the results with input from P.D.K., M.R.B., and W.P.P. F.R. and M.R.B. analyzed the data with input from P.D.K., W.P.P., M.T., and Y.Y. M.R.B. and F.R. wrote the first draft of the manuscript and Supplmenentary Information with significant contributions from M.T., Y.Y., P.D.K., and W.P.P. K.K.B. and F.X.K. provided input and feedback throughout the process.  All authors contributed to the writing and editing of the manuscript.

\section*{Competing Interests}
The authors declare no competing interests.

\clearpage
\setcounter{figure}{0}
\renewcommand{\figurename}{Fig.}
\renewcommand{\thefigure}{S\arabic{figure}}

\title{Supplementary Information for:\\
On-chip sampling of optical fields with attosecond resolution}

\author{\small{Mina R. Bionta$^{1,\dagger,*}$, Felix Ritzkowsky$^{2,\dagger,*}$, Marco Turchetti$^{1,\dagger}$, Yujia Yang$^1$, Dario Cattozzo Mor$^1$, William P. Putnam$^3$, Franz X. Kärtner$^2$,  Karl K. Berggren$^1$, and Phillip~D.~Keathley$^{1,*}$}}
\address{$^1$Research Laboratory of Electronics, Massachusetts Institute of Technology, 77 Massachusetts Avenue, Cambridge, MA 02139, USA\\
$^2$Deutsches Elektronen Synchrotron (DESY) \& Center for Free-Electron Laser Science, Notkestra\ss e 85, 22607 Hamburg, Germany\\
$^3$Department of Electrical and Computer Engineering, University of California, Davis, 1 Shields Ave, Davis, CA 95616, USA\\
$^\dagger$These authors contributed equally to this work.
}
\email{$^*$e-mail: mbionta@mit.edu; felix.ritzkowsky@desy.de; pdkeat2@mit.edu}
\maketitle

\section{Experimental Setup\label{sec:exp_setup}}

\begin{figure}[h]
    \centering
    \includegraphics[width=0.75\textwidth]{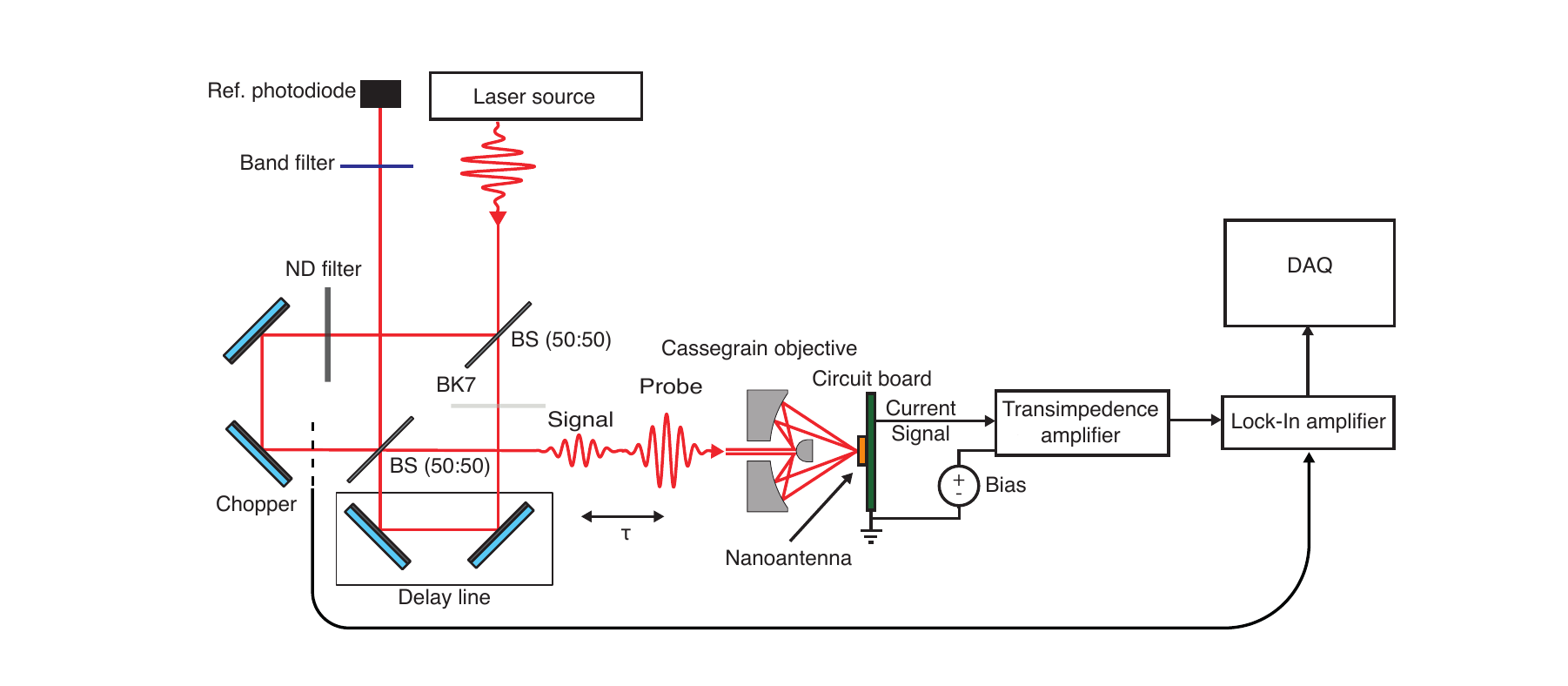}
    \caption{\small\textbf{Experimental Setup}
    Overview of the optical layout and signal detection chain of our experiments. Abbreviations: BS: beamsplitter, ND: neutral density filter, DAQ: data acquisition.
    \label{fig:exp_setup}}
\end{figure}
A CEP-stable, 78 MHz Er:fiber-based supercontinuum laser source was used, with a central wavelength of $\sim$1170~nm and pulse duration of $\sim$10~fs FWHM. A dispersion-balanced Mach-Zehnder interferometer was used to generate the pulse pairs for the experiment (Fig.~\ref{fig:exp_setup}). An Inconel reflective neutral density (ND) filter of optical density (OD) 4 on a 2~mm thick BK7 substrate (Thorlabs) was placed in one arm and used to generate a weak signal pulse with pulse energy of $\sim$5~fJ. An optical chopper was placed in this weak arm for lock-in amplification and detection. The strong, driver arm had a pulse energy of $\sim$50~pJ. A corresponding 2~mm thick BK7 window was placed in the driver arm to balance the dispersion between arms. The added chirp from the glass was precompensated using the prism compressor. The delay between the two pulses was controlled with a home built \SI{15}{\micro\meter} piezo stage. The generated electron emission is collected and amplified by a transimpedance amplifier (FEMTO Messtechnik GmbH). The resulting voltage signal is demodulated by the Lock-In amplifier with the 200~Hz frequency of the chopper wheel and subsequently low-pass filtered.

\section{Discussion of Sampling Bandwidth\label{sec:BW}}

\begin{figure}[h]
    \centering
    \includegraphics{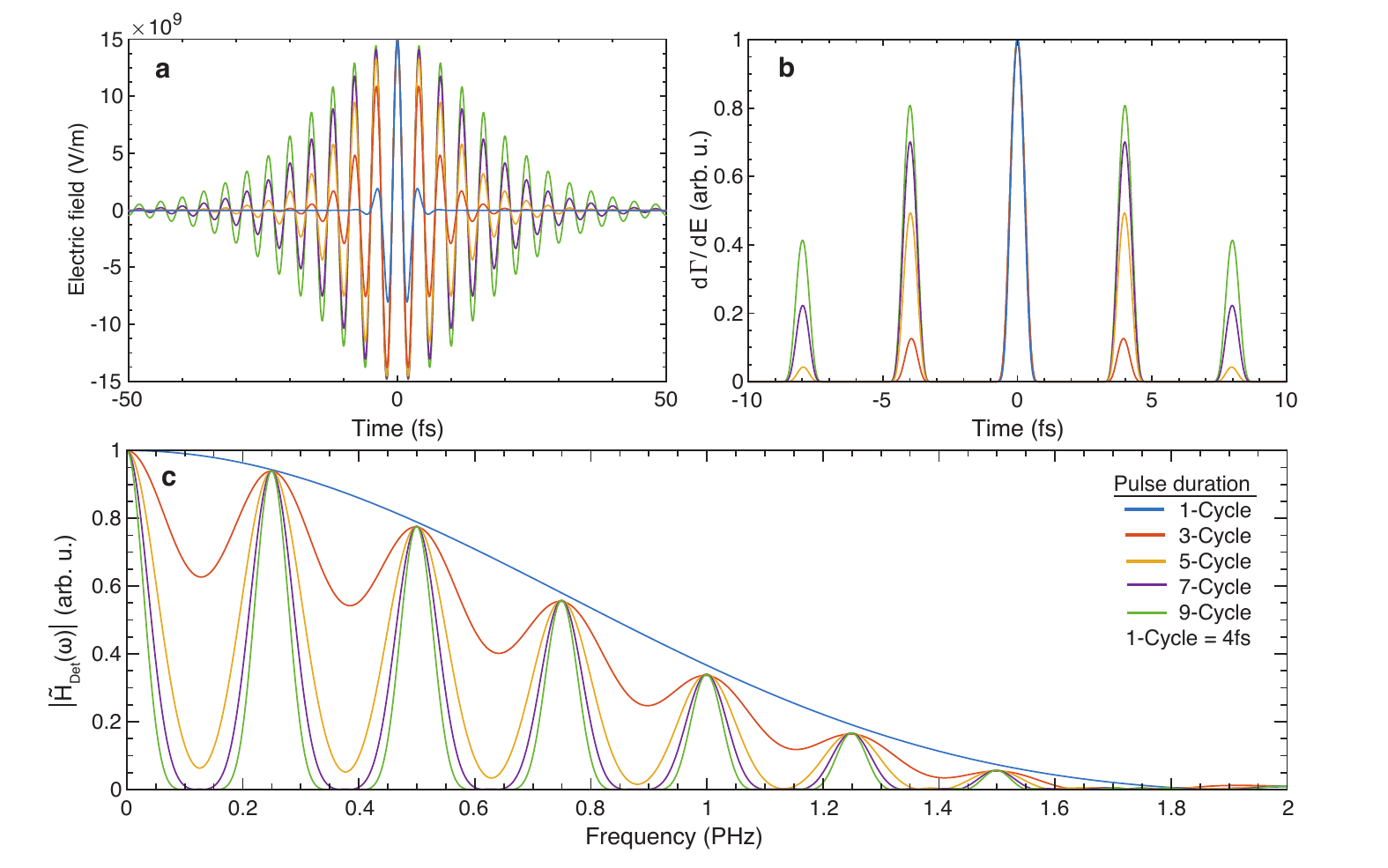}
    \caption{\small\textbf{Sampling bandwidth as a function of pulse duration.}
    \textbf{a,} Electric-field transients for near-infrared pulses with a FWHM duration of 1-, 3-, 5-, 7-, and 9-cycles and a central frequency of $\SI{250}{THz}$.
    \textbf{b,} Calculation of $\left.\dv{\Gamma}{E}\right\vert_{E_\mathrm{D}(t)}$ for the field transients shown in \textbf{a} and assuming $F_t = \SI{78.7}{\volt\per\nano\meter}$ as the characteristic tunneling field.
    \textbf{c,} Fourier transform of $\left.\dv{\Gamma}{E}\right\vert_{E_\mathrm{D}(t)}$ showing the accessible sampling bandwidth provided by the field transients shown in \textbf{a}.
    \label{fig:SamplingBW1}}
\end{figure}

A strong local electric-field transient (driver) drives the electron emission at the metallic nanoantenna \supercite{Rybka2016,Putnam2017,Keathley2019,Ludwig2020}. For simplicity in this section we will be discussing the field driving the emission at a surface, $E_\mathrm{D}(t)$. When a weak electric-field waveform (signal) perturbs the emission process, the detected time-averaged current is proportional to the electric field of the small signal. The small-signal gain, as defined by $\left.\dv{\Gamma}{E}\right\vert_{E_\mathrm{D}(t)}$, is therefore dictated by the strong driving electric field waveform. To demonstrate the influence of the FWHM of the driving pulse duration on the sampling bandwidth, we calculated $\tilde{H}_\text{Det}(\omega)$ for 1-, 3-, 5-, 7-, and 9-cycle sech$^2$ driver pulses each with a central frequency of $\SI{250}{THz}$ and a peak field strength at the antenna surface of \SI{15}{GV m^{-1}} (see Fig.~\ref{fig:SamplingBW1}a). 

The small-signal gain $\left.\dv{\Gamma}{E}\right\vert_{E_\mathrm{D}(t)}$ was calculated by assuming Fowler-Nordheim tunneling emission with a characteristic tunneling field  of $F_t = \SI{78.7}{\volt\per\nano\meter}$. Fig.~\ref{fig:SamplingBW1}b shows the effective gate signal $\left.\dv{\Gamma}{E}\right\vert_{E_\mathrm{D}(t)}$ for the sampling process for each pulse duration. Only the single-cycle pulse (blue) exhibits an isolated peak.  However, for driver pulses with an increasing number of cycles, satellite pulses start to emerge. For the 9-cycle case (green traces) the height of satellite pulses at $\SI{-4}{fs}$ and $\SI{4}{fs}$ approach the height of the center peak. Fig.~\ref{fig:SamplingBW1}c shows the Fourier transform of $\left.\dv{\Gamma}{E}\right\vert_{E_\mathrm{D}(t)}$.

The sampling bandwidth generated by a single-cycle field transient shows a smooth response from DC to \SI{1.8}{PHz} and corresponds to the Fourier transform of the isolated peak in Fig.~\ref{fig:SamplingBW1}b (blue traces). With increasing pulse duration, the bandwidth becomes increasingly modulated due to the destructive interference of the additional peaks in the gate signal. The modulation is periodic with the frequency $f_0$ of the driving electric field at $\SI{250}{THz}$ and exhibits maxima at the higher harmonics $n\cdot f_0$ for $n~\epsilon ~ \mathbb{N}$. We highlight that although a 5-cycle driver waveform results in strong modulation of the sampling response $\tilde{H}_\text{Det}(\omega)$, the sampling response does not completely vanish at the minima (yellow traces). However, for driver pulses having a FWHM duration greater than five cycles, we find that the sampling response completely vanishes at the minima. This sampling technique allows for detection of higher harmonics of the driving signal regardless of the pulse duration, which originates from the fact that the individual peaks are deeply sub-cycle in duration \supercite{Keathley2019}.

\section{Carrier-Envelope Phase Discussion\label{sec:CEP}}

\begin{figure}[h]
    \centering
    \includegraphics{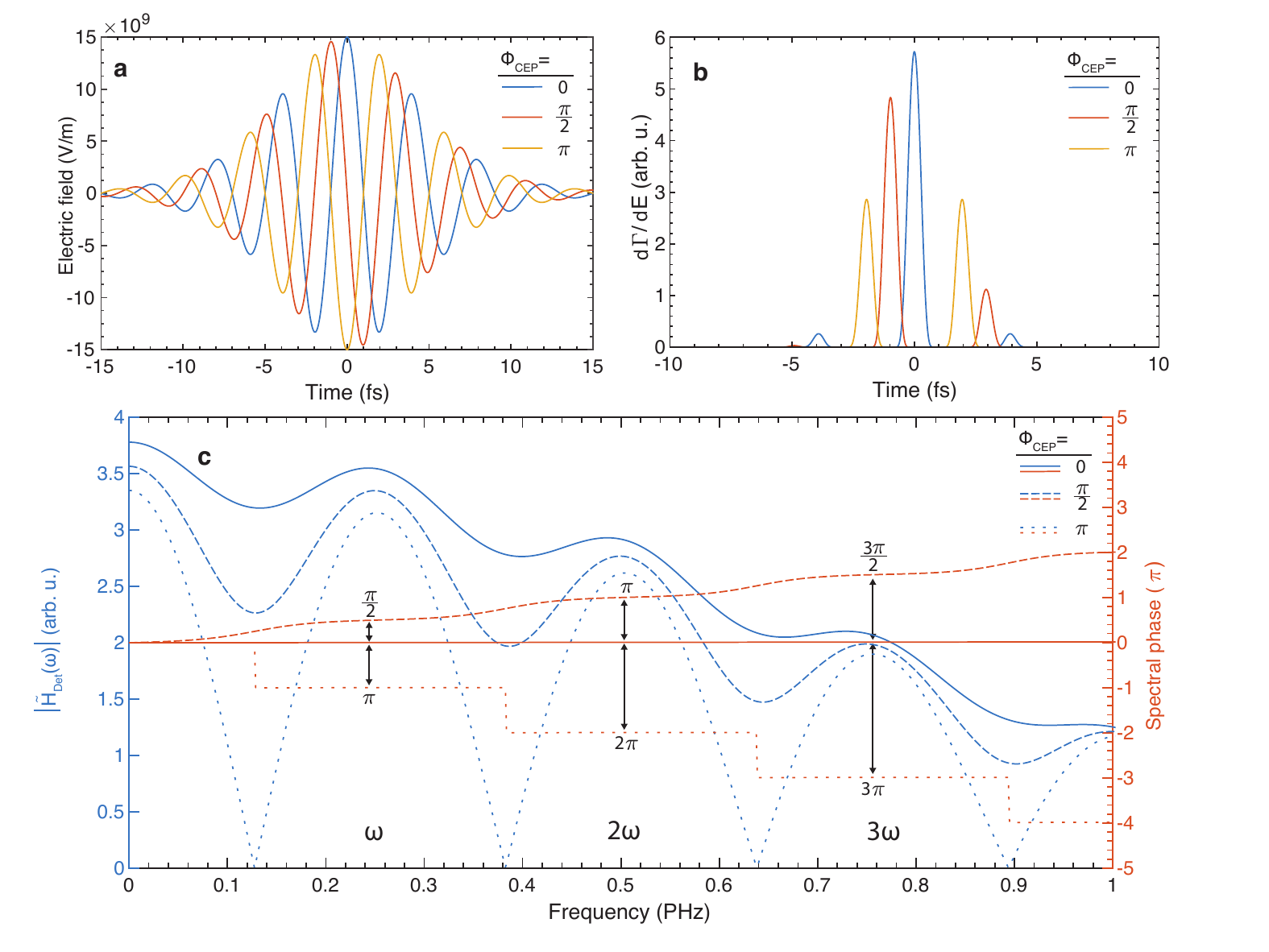}
    \caption{\small\textbf{Sampling response as a function of CEP.}
    \textbf{a,} Calculated sech$^2$ pulse centered at \SI{250}{THz} with a pulse duration of \SI{10}{fs} (2.5 cycle), a peak electric field of \SI{15}{GV m^{-1}}, and a $\Phi_{\mathrm{CEP}} = 0,~\frac{\pi}{2},~\pi$. \textbf{b,} The small signal gain $\left.\dv{\Gamma}{E}\right\vert_{E_\mathrm{D}(t)}$ is calculated by assuming Fowler-Nordheim tunneling emission with a characteristic tunneling field of $F_t = \SI{78.7}{\volt\per\nano\meter}$. The electric-field transients used here correspond to \textbf{a}. \textbf{c,} The spectral amplitude and phase of the complex sampling response of $\tilde{H}_{\mathrm{Det}}(\omega)$ as a function of frequency. Calculated for $\Phi_{\mathrm{CEP}} = 0,~\frac{\pi}{2},\pi$. 
    \label{fig:SamplingCEP}}
\end{figure}

The carrier-envelope phase (CEP) of a few cycle pulse plays a significant role in strong-field physics and heavily influences the electron emission characteristics from resonant nanoantenna devices. In this section we discuss the role of the driving waveform's CEP in the sampling process. For simplicity in this section we will be discussing the field driving the emission at a surface, $E_\mathrm{D}(t)$. 

For our analysis, we calculated the complex sampling response $\tilde{H}_{\mathrm{Det}}(\omega)$ assuming a sech$^2$ driving pulse with a central frequency of \SI{250}{THz} and a pulse duration of \SI{10}{\femto\second} ($\sim$2.5 cycle), as given by the output of the laser used to experimentally verify device performance.  As in Sec.~\ref{sec:BW}, the incident electric field was taken to be \SI{15}{GV/m}.  The results are plotted in Fig.~\ref{fig:SamplingCEP}a for various CEP values of the driving pulse. The small signal gain $\left.\dv{\Gamma}{E}\right\vert_{E_\mathrm{D}(t)}$ was calculated by assuming Fowler-Nordheim tunnel emission with a characteristic tunneling field  of $F_t = \SI{78.7}{\volt\per\nano\meter}$ and is plotted in Fig.~\ref{fig:SamplingCEP}b. In Fig.~\ref{fig:SamplingCEP}c the complex sampling response $\tilde{H}_{\mathrm{Det}}(\omega)$ derived from $\left.\dv{\Gamma}{E}\right\vert_{E_\mathrm{D}(t)}$ is shown. 

The CEP, $\Phi_{\mathrm{CEP}}$, of the driving pulse dictates the amplitude of the modulation of $\tilde{H}_\text{Det}(\omega)$.
For the driver pulse duration modeled in Fig.~\ref{fig:SamplingCEP}a, a cosine shaped pulse ($\Phi_{\mathrm{CEP}}$ = 0) exhibits minimal modulation of the sampling bandwidth, which corresponds to an isolated electron burst with small satellites in the time-domain if the pulse is sufficiently short (see Fig.~\ref{fig:SamplingBW1}b).
A CEP of $\Phi_{\mathrm{CEP}}=\pi$ corresponds to a negative cosine shaped pulse, which corresponds to two electron bursts of equal height, resulting in the sharp minima in the sampling bandwidth as shown in Fig.~\ref{fig:SamplingCEP}c (dotted traces).
More importantly, with an adequately short driving pulse, it is possible to choose an appropriate $\Phi_{\mathrm{CEP}}$ value such that only one electron burst dominates the field emission process, resulting in a smooth, unmodulated $\tilde{H}_{\mathrm{Det}}(\omega)$ from DC to \SI{1}{\peta\hertz}, as shown in Fig.~\ref{fig:SamplingBW1}c. Nevertheless independently of $\Phi_{\mathrm{CEP}}$ a full octave of spectrum can still be sampled with  distortion due to $\tilde{H}_\mathrm{Det}$.

Another important characteristic of the sampling process to consider is the absolute phase of the sampled output. When $\Phi_\mathrm{CEP} = 0$, a dominant electron burst exists in the time domain and the absolute phase of the signal pulse will be transferred to the sampled output, as $\tilde{H}_\mathrm{Det}(\omega)$ will be a purely real function (see Fig.~\ref{fig:SamplingCEP}c). For comparison, if $\Phi_\mathrm{CEP} \neq 0$ the spectral phase of $\tilde{H}_\mathrm{Det}(\omega)$ is not flat. As shown in Fig. \ref{fig:SamplingCEP}, this phase resembles a stair function with plateaus of flat phase around the central frequency $\omega_0$ and its harmonics. Looking closely at Fig. \ref{fig:SamplingCEP}, we see that we can write the spectral phase at the nth harmonic as $\angle \tilde{H}_\mathrm{Det}(n\omega)=n\cdot \Phi_\mathrm{CEP}$ for $n
~\epsilon
~\mathbb{N}$. With these spectral phase behaviors, we then see that the constant phase component of the sampled output becomes the difference between that of the sampling pulse, $n\cdot \Phi_\mathrm{CEP}$, and that of the signal, $\Phi_\mathrm{S}$. Therefore, the constant, or absolute, phase of the sampled output can be written $\Phi_\mathrm{S}-n\cdot \Phi_\mathrm{CEP}$. In the case where the driving pulse, $E_\mathrm{D}$, and the signal pulse, $E_\mathrm{S}$, originate from the same laser source, they will share a common $\Phi_\mathrm{CEP}$, and in this case, the absolute phase of the sampled pulse will therefore be zero. Importantly, we should note that this result is independent of $\Phi_\mathrm{CEP}$, and even laser sources with a carrier envelope offset $f_\mathrm{CEO}\neq 0$ can be used for sampling. Lastly, we should additionally note that in stark contrast to other phase-sensitive techniques, like homo- and hetero-dyne detection, the absolute phase of $E_\mathrm{D}$ can be derived unambiguously \textit{in-situ} from the field emission current generated by $E_\mathrm{D}$ in our devices, as demonstrated in \cite{Rybka2016,Yang2019}. 

\section{Field-Sampling Measurements with 200~nm Devices\label{sec:200nm}}
\begin{figure}[h!]
    \centering
    \includegraphics{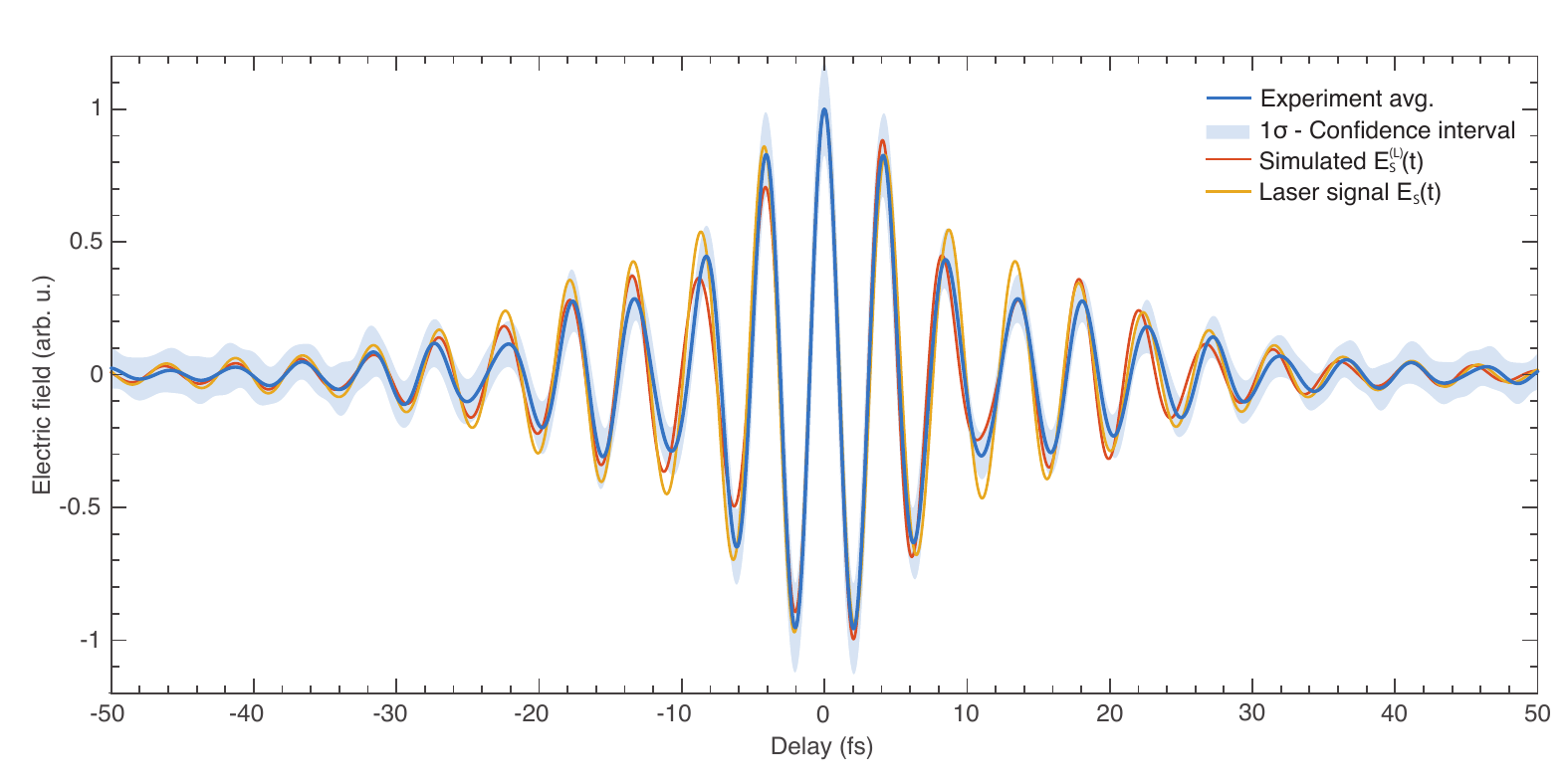}
    \caption{\small\textbf{Experimental field sampling results using 200 nm devices.} Time-domain results for 200~nm devices comparing measured (blue) and simulated (red) near-fields to the calculated incident laser signal (yellow). 
    Here, negative delays indicate the driver pulse arrives before the signal pulse.
    The 200~nm device is designed to be off-resonant with the laser pulse and the measured trace yields good agreement to the calculated laser output.
    The 1$\sigma$-confidence interval is shown as a blue shaded ribbon centered at the average value (blue solid line) retrieved from 47 scans.
    \label{fig:Data_200nm}}
\end{figure}
\begin{figure}[h!]
    \centering
    \includegraphics{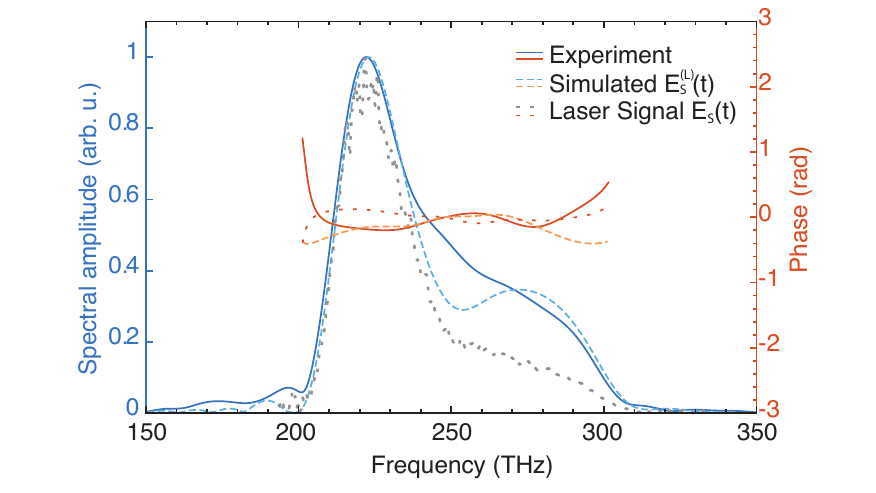}
    \caption{\small\textbf{Frequency-domain analysis of 200 nm device results.} Frequency-domain analysis comparing measured (solid) and simulated (dashed) near-fields for 200~nm devices to the calculated incident laser signal (dotted). 
    The 200~nm device is designed to be off-resonant with the laser pulse, thus the measured and simulated spectrum only show a single spectral peak corresponding with that of the laser spectrum at $\approx220~\mathrm{THz}$.
    \label{fig:Data_200nm_spec}}
\end{figure}

Our technique was also tested using devices consisting of triangular antennas with a 200~nm height.
These devices were designed to be off-resonant with the laser pulse and were fabricated on a separate chip from the 240~nm antenna.
Fig.~\ref{fig:Data_200nm} presents the acquired cross-correlation trace (blue) for these devices.
For each data set, 47 scans of 5 seconds acquisition time over the \SI{100}{fs} time window were performed. Post-processing was done in Matlab.
Each data set was Fourier transformed and windowed from \SI{150}{THz} to \SI{350}{THz} with a tukey-window (steepness of $\alpha=0.2$). The resulting output was averaged in the time-domain.

We find good agreement between the measured trace (blue) to the simulated local signal field, $E_\mathrm{S}^\mathrm{(L)}(t)$ (red). We note that both the measurement and simulated local signal fields are both slightly shorter than the calculated laser output (yellow). The reason for this is apparent when examining the pulses in the frequency domain as shown in Fig.~\ref{fig:Data_200nm_spec}.  
While the main spectral peak at $\approx 220~\mathrm{THz}$ agrees with the measured laser spectrum (gray dotted curve) and the expected antenna response (light blue dashed curve), both the simulated and experimental local signal field spectra exhibit an enhanced shoulder out to \SI{300}{\tera\hertz} relative to the measured laser output spectrum (solid blue curve).  This is due to the plasmonic resonance which enhances these higher frequency components, resulting in a shorter time domain response of the local fields relative to the incident fields after interaction with the antenna.  

\section{Data Processing and Error Analysis}

\begin{figure}[h!]
    \centering
    \includegraphics[width=1\linewidth]{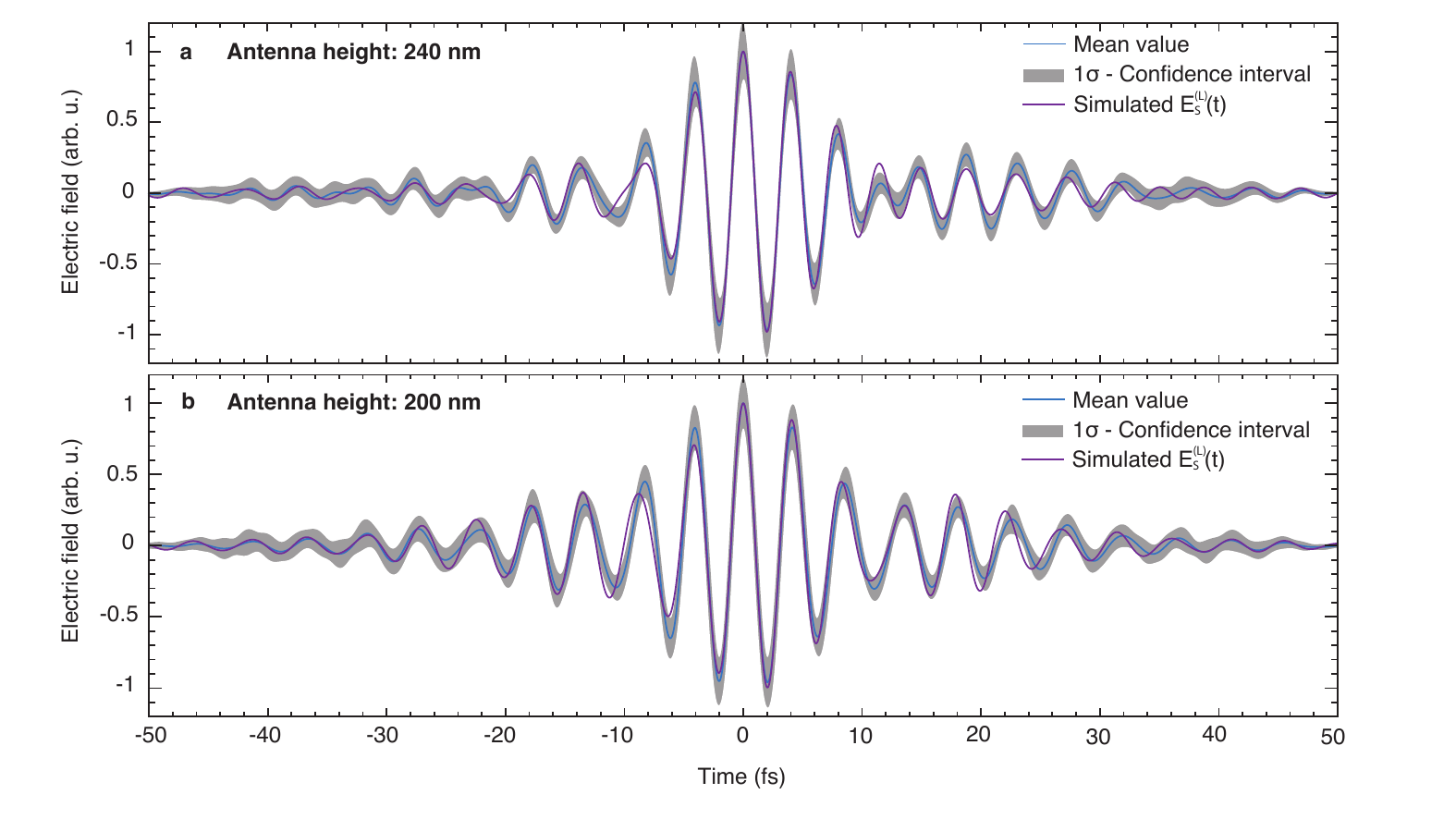}
    \caption{\small\textbf{Mean value and 1$\sigma$-confidence interval} 
    Time-domain measurement and simulation for \textbf{a,} 240~nm devices (Fig.~3, main text) and \textbf{b}, 200~nm devices (Fig.~\ref{fig:Data_200nm}).
    The blue curves shows the mean value for every electric field/time coordinate over all individual scans.
    The grey ribbon shows the 1$\sigma$-confidence interval for the respective coordinate.
    For comparison, the simulated electric field is shown in purple.
    \label{fig:conf}}
\end{figure}

To determine the error in our measurement, we took the Fourier transform of the each of the $\sim$50 individual data sets and applied a tukey-window in the frequency-domain with a steepness of $\alpha = 0.2$ from \SI{150}{THz} to \SI{350}{THz}.
The windowed data sets were then back transformed into the time-domain and averaged for each time coordinate over all data sets.
To determine the $1\sigma$-confidence interval the standard deviation was calculated for each time coordinate over all data sets.
The result is shown in Fig. \ref{fig:conf} and compared to the respective simulation shown in Fig.~3 (main text) for the 240~nm devices and Fig. \ref{fig:Data_200nm} for the 200~nm devices. 

\section{Source Spectral Phase Measurements\label{sec:2DSI}}
\begin{figure}[h!]
    \centering
    \includegraphics[width=1\linewidth]{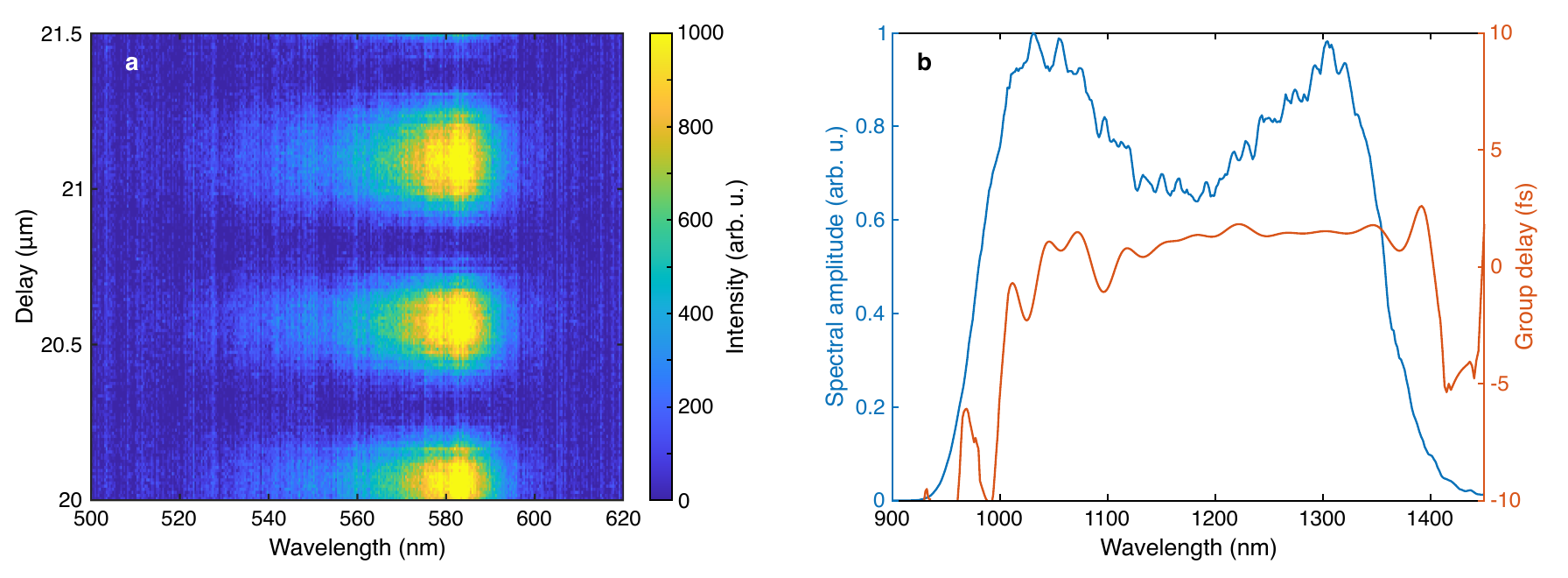}
    \caption{\small\textbf{Source spectral phase characterization using 2DSI.} \textbf{a,}
    Raw 2DSI spectrogram of the source in the experiment conditions. \textbf{b,} Retrieved group delay (red) and laser spectrum (blue). The optimized values of shear frequency and upconversion wavelength are $f_\text{shear} = 5.5$~THz and $\lambda_\text{up} = 1050$~nm.
    \label{fig:2DSI}}
\end{figure}

In order to characterize the spectral phase of our supercontinuum source we performed two-dimensional spectral shearing interferometry (2DSI) measurements~\supercite{Birge2006}.
Two spectrograms were obtained for the measurement: the first with the laser in similar conditions to that of the experiment, and the second with an added 1.5~mm fused silica window placed in the beam path.
The spectrogram of the source in the experimental conditions is shown in Fig. \ref{fig:2DSI}a. 
The second spectrogram taken with an additional propagation through 1.5~mm fused silica was used to calibrate the shear frequency $f_\text{shear}$ and upconversion wavelength $\lambda_\text{up}$ needed for group delay retrieval from the 2DSI measurement.  Using an optimization routine, we found the values for $f_\text{shear}$ and $\lambda_\text{up}$ that resulted in the minimum error between the group delay difference measured with and without the fused silica using 2DSI and that predicted using the known optical properties of fused silica.    
The resulting retrieved group delay and the spectrum of our laser source are reported in Fig. \ref{fig:2DSI}b.
\end{document}